\documentclass[amsmath,amssymb,aps,prl,superscriptaddress,10pt,reprint,longbibliography]{revtex4-1}

\usepackage{graphicx}

\usepackage{textcomp} 

\usepackage[colorlinks,linkcolor=blue,citecolor=blue,urlcolor=blue]{hyperref}

\def\bg#1\eg{\begin{align}#1\end{align}}

\begin{document}

\title{The influence of pump coherence on the generation of\\
position-momentum entanglement in down-conversion}

\author{Wuhong Zhang}
\affiliation{Department of Physics, Jiujiang Research Institute and Collaborative Innovation Center for Optoelectronic Semiconductors and Efficient Devices, Xiamen University, Xiamen 361005, China}
\affiliation{Department of Physics, University of Ottawa, 25 Templeton Street, Ottawa, Ontario K1N 6N5, Canada}
\author{Robert Fickler}
\affiliation{Department of Physics, University of Ottawa, 25 Templeton Street, Ottawa, Ontario K1N 6N5, Canada}
\affiliation{current address: Institute for Quantum Optics and Quantum Information (IQOQI),
Austrian Academy of Sciences, Boltzmanngasse 3, 1090 Vienna, Austria}
\author{Enno Giese}
\email{enno.a.giese@gmail.com}
\affiliation{Department of Physics, University of Ottawa, 25 Templeton Street, Ottawa, Ontario K1N 6N5, Canada}
\affiliation{current address: Institut f\"ur Quantenphysik and Center for Integrated Quantum Science and Technology (IQ\textsuperscript{ST}), Universit\"at Ulm, Albert-Einstein-Allee 11, D-89081, Germany}
\author{Lixiang Chen}
\email{chenlx@xmu.edu.cn}
\affiliation{Department of Physics, Jiujiang Research Institute and Collaborative Innovation Center for Optoelectronic Semiconductors and Efficient Devices, Xiamen University, Xiamen 361005, China}
\author{Robert W. Boyd}
\affiliation{Department of Physics, University of Ottawa, 25 Templeton Street, Ottawa, Ontario K1N 6N5, Canada}
\affiliation{Institute of Optics, University of Rochester, Rochester, NY 14627, United States of America}

\maketitle

\textbf{Strong correlations in two conjugate variables are \emph{the} signature of quantum entanglement and have played a key role in the development of modern physics~\cite{Einstein35,Reid09}.
Entangled photons have become a standard tool in quantum information \cite{flamini2018photonic} and foundations \cite{Shalm15,Giustina15}.  
An impressive example is position-momentum entanglement of photon pairs~\cite{Howell04}, explained heuristically through the correlations implied by a common birth zone and momentum conservation. However, these arguments entirely neglect the importance of the `quantumness', i.e. coherence, of the driving force behind the generation mechanism. We study theoretically and experimentally how the correlations depend on the coherence of the pump of nonlinear down-conversion. In the extreme case - a truly incoherent pump - only position correlations exist. By increasing the pump's coherence, correlations in momenta emerge until their strength is sufficient to produce entanglement. Our results shed light on entanglement generation and can be applied to adjust the entanglement for quantum information applications.}

Entanglement of photons has been explored among different degrees of freedom, such as polarization~\cite{Freedman72,Shalm15,Giustina15}, time and frequency~\cite{Franson89,Kwiat93}, position and momentum~\cite{Howell04} as well as angular position and orbital angular momentum~\cite{Mair01,Leach10}. 
Entanglement of two-dimensional systems, in analogy to classical bits, is \emph{the} primary resource for quantum communication and processing \cite{flamini2018photonic}.
In addition, multiple-level quantum systems can show high-dimensional entanglement with a~high complexity~\cite{Krenn14,Xie15,Wang18} and can be exploited for various quantum information tasks~\cite{Erhard18}.
Position-momentum entanglement as a continuous degree of freedom is the ultimate limit of high-dimensional entanglement and its deeper understanding
is essential for the development of novel quantum technologies.

\begin{figure*}[t]
\centering
\includegraphics[width=1.7\columnwidth]{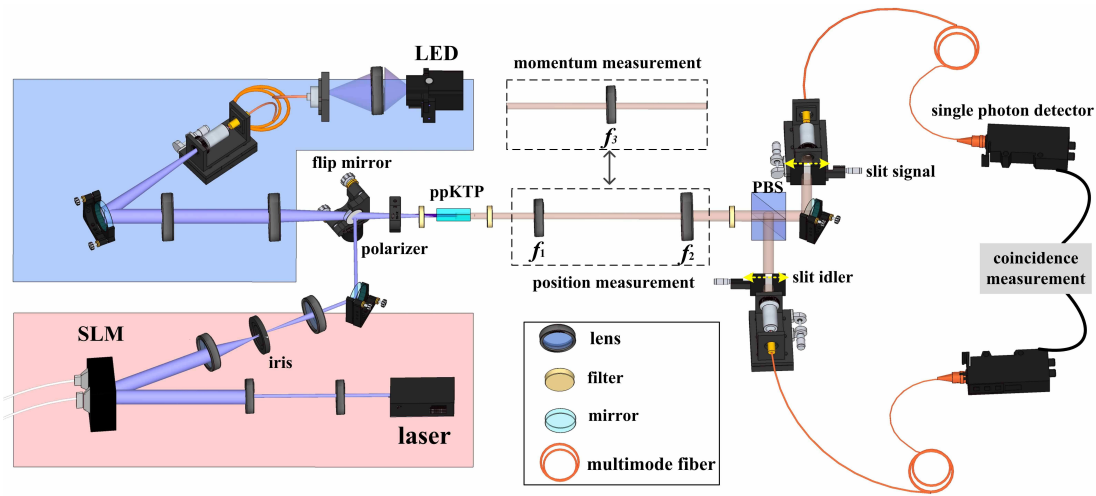}

\caption{
Schematic of the experimental setup.
Photon pairs are generated by pumping a~type-II nonlinear crystal (ppKTP) with either a~laser beam with adjustable transverse coherence (red shaded beam path) or a~beam derived from an LED that is spatially incoherent (blue shaded beam path).
The coherence of the laser is tuned by modulating the transverse phase profile with a spatial light modulator (SLM).
A polarizing beam splitter (PBS) splits the pairs and their joint spatial distributions are measured by independently movable slits in each arm.
They are followed by bucket detector systems consisting of microscope objectives, multimode fibers, single-photon detectors.
Position correlations are registered by a~coincidence measurement in the imaging plane ($f_1$ and $f_2$), while momentum correlations are observed in the focal plane of a~lens $f_3$ (Fourier transform plane, or momentum space).
}\label{fig_setup}
\end{figure*}

Position-momentum-entangled photon pairs can be rather straight-forwardly generated in spontaneous parametric down-conversion (SPDC)~\cite{Howell04,DAngelo04}, the workhorse of many quantum optics labs.
In this process, a~strong pump beam spontaneously generates a~pair of signal and idler photons through a~nonlinear interaction.
Formation of position-momentum entanglement is often explained by simple heuristic arguments:
A pump photon is converted at one particular transverse position into signal and idler.
Due to this common birth place, they are correlated in position. In addition, transverse momentum conservation requires the generated photons to travel in opposite directions, i.e. they are anti-correlated in momentum. Hence, in an idealized situation the generated pairs can be perfectly correlated in both, the position and momenta, which is the key signature of quantum entanglement.
However, these arguments have not taken the coherence properties of the pump beam, i.e. the quantum aspect of the driving force behind the pair generation, into account. In this letter, we study how the generation of position-momentum entangled photon pairs relies on the coherence properties of the pump.
For that, we pump a nonlinear crystal by a coherent light source (a laser), a \emph{true} incoherent source (an LED), and examine the transition between these extreme cases by pumping with pseudo-thermal light of variable partial coherence.
We find that the strength of the momentum anti-correlation depends strongly on the coherence of the pump so that the degree of entanglement can be adjusted.
Fundamentally, our analysis demonstrates that the lack of momentum correlation does not imply an violation of the conservation of momenta; it shows that the coherence of the pump, i.e. its `quantumness', is crucial for the generation of entangled photons. 

A~first theoretical analysis~\cite{Jha10,Giese18} of the pair-generation process shows that the angular profile of the pump and its coherence is transferred to the down-converted light. Thus, it determines the uncertainty of the anti-correlation and also effects the generation of entanglement.
Along similar lines, the influence of different coherent pump profiles on entanglement and on the propagation of the generated pairs have been already explored~\cite{Monken98,Law04,Chan07,Gomes09,Walborn10}. The impact of temporal coherence of the pump has been investigated in~\cite{Burlakov01,Jha08,Kulkarni17}.

Our experimental setup (see Fig.~\ref{fig_setup}) is designed in a~flexible manner so that switching between the the laser and the LED (red and blue shaded regions in Fig.~\ref{fig_setup}) can be easily accomplished with a~flip mirror. We can further change between detecting position and momentum correlations simply by using a different set of lenses (see more details in Methods).
To investigate entanglement, we measure the probability distributions of the distance $x_-\equiv (x_s-x_i)/\sqrt{2} $ between singal ($s$) and idler ($i$) photons, as well as their average momentum $p_+ \equiv (p_s+p_i)/\sqrt{2}$ and compare the results obtained for both sources.
A high correlation in the positions and momenta reflects itself in small uncertainties $\Delta x_-^2$ and $\Delta p_+^2$.
In fact, they are often used to verify entanglement of continuous variables, as it is possible that the product of the uncertainties \emph{violates} the inequality~\cite{Reid09,Schneeloch16}
\bg\label{e_separability}
\Delta x_-^2 \Delta p_+^2 \geq \hbar^2 /4.
\eg

\begin{figure}[t]
\centering
\includegraphics[width=\columnwidth]{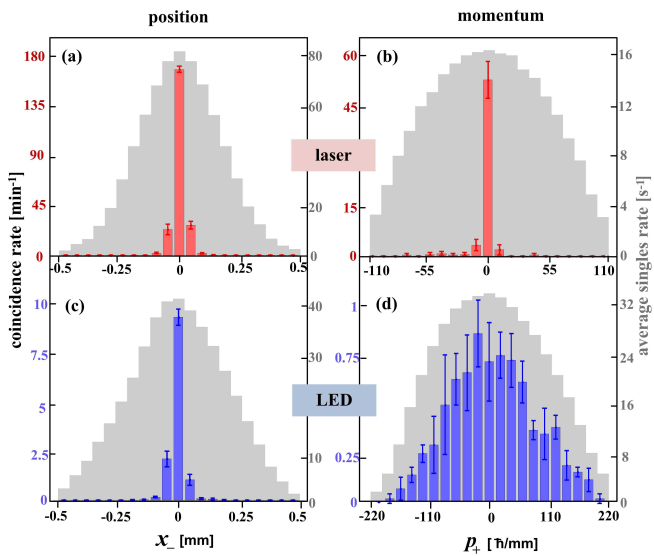}
\caption{
Position correlations and momentum anti-correlations of SPDC pumped by a~laser and an LED.
We obtain the coincidence rates in (a,c) by moving the two slits in opposite directions in the near field (measuring the distribution of $x_-$) and in (b,c) by moving them in the same directions in the far field (measuring the distribution of $p_+$).
The acquisition time of each data point is 1\,min for the laser pump (a,b) and 15\,min for the LED pump (c,d); the error bars are obtained by averaging over five such measurements. To demonstrate the correlation strength graphically, we show the average rate of singles counts (gray distributions in the back; scale on the right side of each plot).
}
\label{fig_entanglement}
\end{figure}

The distributions of $x_-$ for both sources are shown in Fig.~\ref{fig_entanglement}(a) and~(c).
The positions of signal and idler photons are highly correlated and the shapes of the two distributions coincide, underlining argument of a~common birth zone.
For the momenta, the distributions of $p_+$ obtained with a~laser and with an LED differ significantly, see Fig.~\ref{fig_entanglement}(b) and~(d).
The momenta of the photons generated by the laser are anti-correlated, in agreement with the argument of momentum conservation.
We further verify entanglement, since the measured uncertainty product
\bg\label{e_product_laser}
\left.\Delta x_-^2 \Delta p_+^2\right|_\text{laser} = (0.0112 \pm 0.0005 ) \hbar^2
\eg
violates inequality~\eqref{e_separability}.
Here, as well as in all following discussions, we obtain the uncertainties by a~Gaussian fit to the experimental data.
In contrast, the momenta obtained from an LED-pumped source are uncorrelated, and the broad distribution leads to
\bg\label{e_product_LED}
\left.\Delta x_-^2 \Delta p_+^2\right|_\text{LED} = (4.62 \pm 0.93 ) \hbar^2 ,
\eg
consistent with inequality~\eqref{e_separability}, implying that entanglement is not present and seemingly in contrast to the argument of momentum conservation.

\begin{figure}[t]
\centering
\includegraphics[width=\columnwidth]{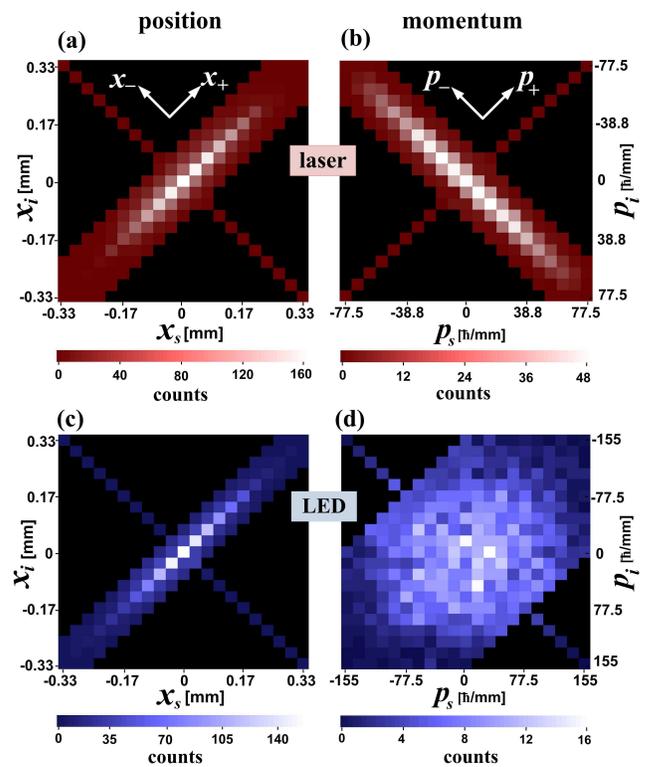}
\caption{
Effect of coherence on joint probability distributions of generated photon pairs.
Parts~(a) and~(b) show the joint position and momentum distributions when pumped with a~laser (red), parts~(c) and~(d) show the respective distributions when pumped with an LED (blue).
Horizontal axes denote the position or momentum of the signal; vertical axes denote the position or momentum of the idler.
The joint distributions show the number of coincidence counts accumulated in 1\,min for the laser (a, b) and in 15\,min for the LED (c, d).
The pairs are strongly correlated in position when pumped with either the laser or the LED, while strong anti-correlation of their momenta occurs only for a~transverse coherent pump beam.
The black areas depict parts that have not been measured, since nearly no counts were expected.
}
\label{fig_distributions}
\end{figure}

For a more detailed analysis, we measure the entire joint probability distributions for position space $\mathcal{P}(x_s,x_i)$ and momentum space $P(p_s,p_i)$ for both the laser and the LED. The results are illustrated in Fig.~\ref{fig_distributions}.
The joint momentum distribution consists two contributions: the angular profile of the pump along the diagonal
and the phase-matching function along the anti-diagonal of $(p_s,p_i)$-space, given by $p_\pm = (p_s \pm p_i)/\sqrt{2}$, respectively (see Methods for more details).
In position space, the distribution has the same structure and can be written as the product of two contributions that can be associated with the spatial profile and the Fourier transform of the phase-matching function along the digaonal and anti-diagonal of $(x_s,x_i)$-space, given by $x_\pm = (x_s \pm x_i)/\sqrt{2}$.

The distributions for a~laser pump are shown in Fig.~\ref{fig_distributions}(a,b).
We observe narrow ellipses along the diagonal in position space ($\Delta x_-/\Delta x_+ = 0.153 \pm 0.003$) and along the anti-diagonal in momentum space ($\Delta p_+/ \Delta p _- = 0.083\pm 0.004 $), which underlines the high degree of position correlation and momentum anti-correlation.
The combination of the two is a~signature of entanglement and these measurements underline our heuristic arguments of a common birth zone and momentum conservation.

The joint position distribution for the LED pump is shown in Fig.~\ref{fig_distributions}(c).
Since we designed the experiment such that the width of the intensity distribution of the LED light in the crystal is comparable to that of the laser, the two distributions are very similar.
We observe a narrow ellipse along the diagonal in position space, i.\,e. the photon pairs are strongly correlated in position ($\Delta x_-/\Delta x_+ = 0.174 \pm 0.003$).
In contrast, the joint momentum distribution for the LED shown in Fig.~\ref{fig_distributions}(d) demonstrates that the two momenta are \emph{uncorrelated} ($\Delta p_+ /\Delta p_- = 1.0 \pm 0.1$).
Because entanglement requires a~strong degree of correlation in both positions \emph{and} momenta, we observe no position-momentum entanglement of photon pairs generated by the LED.
The anti-correlations vanish not because transverse momentum conservation becomes invalid, but because the angular profile of a~transverse incoherent beam is dramatically different from that of a~coherent beam.

\begin{figure}[t]
\centering
\includegraphics[width=\columnwidth]{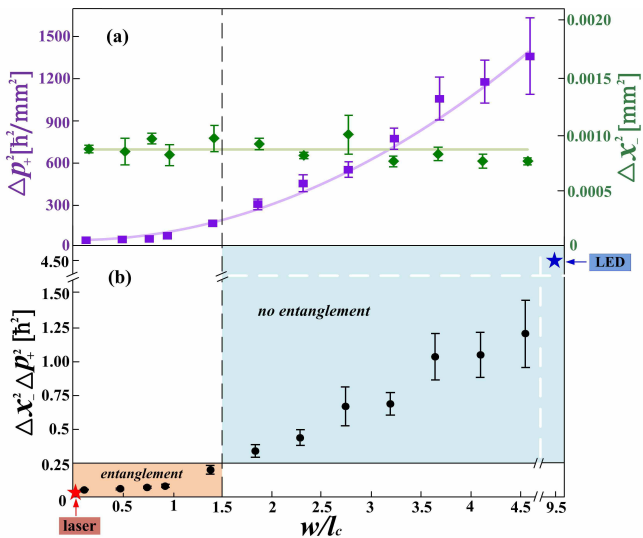}
\caption{
Momentum anti-correlation, position correlation, and the entanglement criterion for pseudo-thermal pump beams with different coherence lengths.
Part (a) shows that $\Delta x_-^2$ (green) is independent of the coherence length, whereas $\Delta p_+^2$ (purple) follows equation~\eqref{e_Delta_p_+}, as highlighted by the fit.
The product $\Delta x_-^2 \Delta p_+^2$ in part (b) increases for decreasing coherence and therefore makes a transition from entangled to classically correlated photon pairs.
The red star represents this product for the (coherent) laser and the blue star represents this product for the (incoherent) LED whose coherence length has been extrapolated from a fit.
}
\label{fig_partially_coherent}
\end{figure}

We complete our study by experimentally invetsigating the effect of the coherence length $l_c$ of a partially coherent beam on the entanglement.
We spatially modulate the laser to generate a pseudo-thermal field that can be described by a Gaussian Schell-model beam~\cite{Mandel95}.
Such a~pump beam with a~beam waist $w$, a~radius of curvature $R$, and a~wave number $k_p$ leads to the variance~\cite{Giese18}
\bg \label{e_Delta_p_+}
\Delta p_+^2 = \hbar^2 / (8w^2) + \hbar^2 w^2 k_p^2 /(2 R^2) + \hbar^2 / (2 l_c^2)
\eg
of the angular profile.
The coherence length $l_c$ causes a~spread similar to the one caused by a~finite radius of curvature $R$.
We tune the coherence length~\cite{Shirai05} through the modulation strength of different random phases imprinted on the pump laser and averaged over 300 patterns (see Methods for more details).
The measured uncertainties $\Delta x_-^2$ and $\Delta p_+^2$ are shown in Fig.~\ref{fig_partially_coherent}(a).
The position correlation remains unchanged and is independent of the coherence length~\cite{Giese18}.
In contrast, the uncertainty $\Delta p_+^2$ scales quadratically with the parameter $w/l_c$, following equation~\eqref{e_Delta_p_+}.
The product $\Delta x_-^2 \Delta p_+^2$ shown in Fig.~\ref{fig_partially_coherent}(b) highlights the impact of $l_c$ on entanglement.  
For sufficiently large coherence (small $w/l_c$), the product is below the bound of $\hbar^2/4$.
For a decreasing coherence length (increasing $w/l_c$), we exceed this bound and cannot verify entanglement.
The laser result from equation~\eqref{e_product_laser} is consistent with the limit of a~fully coherent beam.
The result for the LED from equation~\eqref{e_product_LED} is far beyond what we observed for pseudo-thermal light.
Although an extrapolation from our data would lead to a~rough estimate of 12\,\textmu m for the coherence length of the LED, we emphasize that the Gaussian Schell model does not describe such a~source very well.
We believe that the uncertainty $\Delta p_+$ of the LED is not determined solely by the inverse of $l_c$, but is in addition limited by the finite aperture of the microscope lens, the low pump efficiency and the non-paraxiality of the incoherent light.
An indication of similar effects might be the small difference of $\Delta p_-$ between the laser and LED measurements, which could be caused by the strong focusing of the LED inside the crystal and its small longitudinal coherence~\cite{Chan07,DiLorenzoPires11}.

In summary, we have studied the importance of spatial coherence of the pump to generate position-momentum entangled photons and demonstrated the ability to control the degree of entanglement by tuning the coherence of the pump. 
Since partially coherent beams have been shown to be less susceptible to atmospheric turbulence~\cite{Gbur14}, our configuration might be useful for future long-distance quantum experiments and could offer a testbed for entanglement purification and distillation protocols~\cite{Hage08}.
We have demonstrated that only for idealized situations, i.e. a perfectly coherent pump, the heuristic arguments to explain position-momentum entanglement remain valid, and we have shed light on important subtleties of the underlying phenomena of entanglement.
Our results underline the relevance of the coherence of the driving force for the generation of entanglement, not only in quantum optics but also in other physical systems such as matter waves or Bose-Einstein condensates.

\small
\section{methods}
\textbf{Experimental Setup}: In our experiment, the coherent pump source is a~laser diode module (Roithner LaserTechnik, RLDE405M-20-5), which can be turned into a pseudo-thermal light source by modulating the transverse phase profile with a~spatial light modulator (SLM, Hamamatsu X10468-05).
The SLM is either used as a simple mirror or to generate a pseudo-thermal light source with varying transverse coherence~\cite{Shirai05} and a~beam waist of $w=0.11$\,mm in the crystal.
The incoherent source is a~blue LED~\cite{Tamosauskas10,Galinis11} with a~center wavelength of 405\,nm and an output power of up to 980\,mW (Thorlabs M405L3).
To ensure a~Gaussian-like beam profile while maintaining transverse incoherence, we couple the light into a~400-\textmu m-core multimode fiber.
The out-coupled LED beam is then demagnified by a~$4f$-system before it enters the crystal.
To ensure the same polarization for both sources, we introduce polarizers in both beam paths.
We additionally add a~3-nm-bandpass filter at 405\,nm in front of the crystal to reduce the broad spectrum of the LED.
After this filtering, we measure a~pump power of 20\,\textmu W for the laser and 130\,\textmu W for the LED at the crystal.

In all pump scenarios, the photon pairs are generated by a~1\,mm$\times$2\,mm$\times$5\,mm periodically poled potassium titanyl phosphate crystal (ppKTP), which is phase-matched for type-II collinear emission.
A~long-pass filter and a~3-nm-spectral filter at 810\,nm after the crystal block the pump beam and ensure that only frequency-degenerate photons are detected.
We split the photon pairs into two separate paths by means of a~polarizing beam splitter.
In each path we place a~narrow vertical slit of about 100\,\textmu m width, which can be translated in the horizontal direction and detects either position or momentum depending on the optical system (see below).
Photons passing through the vertical slits are collected by microscope objectives, coupled into multimode fibers, and detected by avalanche photodiode single-photon counting modules.
The photon coincidence count rate is recorded with a~coincidence window of 1\,ns and as a~function of the two distances $d_s$ and $d_i$ of the slits from the optical axis.
To measure the joint position distribution, we image the exit face of the crystal onto the planes of the slits with a~$4f$-system consisting of two lenses with focal lengths $f_1 = 50$\,mm and $f_2=150$\,mm (placed prior to the beam splitter).
We magnify the down-converted beam to reduce errors that arise from the finite precision of the slit widths.
By replacing the $4f$-system with a~single lens $f_3$ and placing the two slits in the Fourier planes of the lens, we measure correlations of the transverse momenta of the photons.
We use a~focal length of $f_3 = 100$\,mm for the laser and a~shorter focal length of $f_3 = 50$\,mm for the LED to account for the broader momentum distribution of the LED beam.
Again, we record the coincidence count rate as a~function of the position of each slit, and we transform the distance $d_{s,i}$  to momentum through the relation $p_{s,i} \cong \hbar d_{s,i} k_{s,i} / f_3 $.
Here, $k_{s,i}$ denotes the wave number of the signal or idler field.

To generate a Gaussian Schell model beam, we imprint with the SLM different random phase patterns on the pump laser.
The statistics of these random patterns is Gaussian with a transverse width in the crystal of $\delta_\phi=0.11\,$mm.
To tune the coherence length, we vary the strength of the modulation $\phi_0$ and obtain the coherence length from $l_c = \delta_\phi/\phi_0$~\cite{Shirai05}.
For each modulation strength, we display around 300 different patterns, average over the observed counts per measurement setting, and evaluate the obtained uncertainties $\Delta x_-^2$ and $\Delta p_+^2$.

\textbf{SPDC theory}:
In a spontaneous parametric down conversion process, the joint momentum distribution $P(p_s,p_i)=P_\mathcal{E}P_\chi$ consists two parts:
(i) the angular profile of the pump $P_\mathcal{E}\propto |\mathcal{E}(p_s+p_i)|^2$, where $\mathcal{E}$ is the angular field amplitude, 
and (ii) the phase-matching function $P_\chi$, which depends on the the mismatch $\Delta \kappa \equiv \kappa_p-\kappa_s-\kappa_i$.
Here, $\kappa_{p,s,i}$ are the longitudinal components of the wave vectors of the pump, signal, and idler fields.
For a bulk crystal of length $L$, the phase-matching function takes the familiar form $P_\chi\propto \operatorname{sinc}^2(\Delta \kappa L/2)$, but for other configurations it depends on the crystal poling and other properties that arise from the propagation of the light through the medium.
If we assume a~crystal of infinite transverse size, we obtain precise transverse momentum conservation, as is apparent from the argument $p_s+p_i$ of $\mathcal{E}$.
In the paraxial approximation, $\Delta \kappa$ scales as the square of the difference in the transverse momenta $p_s-p_i $, as can be seen from a Taylor expansion of $\kappa_j = \left(k_j^2 -  p_j^2/\hbar^2\right)^{1/2}$ for $p_j\ll \hbar k_j$, where $k_j$ is the modulus of the wave vector of the respective field~\cite{Giese18}.
With the help of a rotated coordinate system $p_\pm \equiv (p_s\pm p_i)/\sqrt{2}$, we can rewrite the angular intensity profile to $P_\mathcal{E}=P_\mathcal{E}(p_+)$ as well as the phase-matching function $P_\chi=P_\chi(p_-)$ such that they are only functions $p_+$ and $p_-$, respectively. After transforming to position space with a Fourier transformation and after an analogue rotation of the coordinates system $x_\pm\equiv (x_s\pm x_i)/\sqrt{2}$, we find a similar structure $\mathcal{P}(x_s,x_i) = \mathcal{P}_\mathcal{E}(x_+) \mathcal{P}_\chi (x_-)$. 
Here, the function $\mathcal{P}_\mathcal{E}(x_+)$ along the diagonal of $(x_s,x_i)$-space corresponds to the intensity profile of the laser and the function $\mathcal{P}_\chi(x_-)$ along the anti-diagonal is connected to the phase-matching function through a~Fourier transformation.

\section{acknowledgements}

We thank Armin Hochrainer for stimulating discussions.
WZ acknowledges the financial support of the China Scholarship Council (CSC).
EG, RF, and RWB are thankful for the support by the Canada First Research Excellence Fund award on Transformative Quantum Technologies and by the Natural Sciences and Engineering Council of Canada (NSERC).
RF acknowledges the financial support of the Banting postdoctoral fellowship of the NSERC.
LC thanks the National Natural Science Foundation of China (11474238, 91636109), the Fundamental Research Funds for the Central Universities at Xiamen University (20720160040), the Natural Science Foundation of Fujian Province of China for Distinguished Young Scientists (2015J06002), and the program for New Century Excellent Talents in University of China (NCET-13-0495).

\section{Contributions}
E.G. and R.F. conceived the idea, E.G. developed the theory, W.Z. and R.F. designed the experiment, W.Z. performed the experiment, W.Z., E.G. and R.F. analyzed the data and wrote the manuscript, L.C. and R.W.B. supervised the project. All authors contributed to scientific discussions.

\bibliography{myref}
\end{document}